\documentclass[english, prx, showkeys, showpacs, reprint, superscriptaddress]{revtex4-1}
\usepackage[latin9]{inputenc}
\usepackage{color}
\usepackage{babel}
\usepackage{verbatim}
\usepackage{bm}
\usepackage{amsmath}
\usepackage{amssymb, mathtools}
\usepackage{graphicx}
\usepackage[unicode=true,pdfusetitle,
 bookmarks=true,bookmarksnumbered=false,bookmarksopen=false,
 breaklinks=false,pdfborder={0 0 0},pdfborderstyle={},backref=false,colorlinks=true]
 {hyperref}
\hypersetup{
 citecolor=blue}

\makeatletter
\usepackage{epstopdf}
\usepackage{amsfonts}
\usepackage{mathrsfs}\usepackage{bm}\usepackage{appendix}
\usepackage{txfonts}
\usepackage{float}
\usepackage{epstopdf}
\usepackage{ulem}

\makeatother

\begin{document}

\title{Cooper Instability and Superconductivity of the Penrose Lattice}
\author{Yu-Bo Liu}
\thanks{These two authors contributed equally to this work.}
\affiliation{School of Physics, Beijing Institute of Technology, Beijing 100081, China}
\author{Juan-Juan Hao}
\thanks{These two authors contributed equally to this work.}
\affiliation{School of Physics, Beijing Institute of Technology, Beijing 100081, China}
\author{Yongyou Zhang}
\email{yyzhang@bit.edu.cn}
\affiliation{School of Physics, Beijing Institute of Technology, Beijing 100081, China}
\author{Ye Cao}
\affiliation{School of Physics, Beijing Institute of Technology, Beijing 100081, China}
\author{Wei-Qiang Chen}
\affiliation{Department of Physics, Southern University of Science and Technology, Shenzhen, 518055, China}
\affiliation{Shenzhen Institute for Quantum Science and Engineering,Southern University of Science and Technology, Shenzhen 518055, China}
\affiliation{Shenzhen Key Laboratory of Advanced Quantum Functional Materials and Devices, Southern University of Science and Technology, Shenzhen 518055, China}
\author{Fan Yang}
\email{yangfan\_blg@bit.edu.cn}
\affiliation{School of Physics, Beijing Institute of Technology, Beijing 100081, China}

\date{\today}
\begin{abstract}
Bulk superconductivity (SC) has recently been observed in the Al-Zn-Mg quasicrystal (QC). To settle several fundamental issues of the SC on the QC, we use an attractive Hubbard model to perform a systematic study on the Penrose lattice. The first issue is the Cooper instability of the QC, i.e., no Fermi surface under an infinitesimal attractive interaction. Starting from the two-electron problem outside the filled Fermi-sea, we analytically prove that an infinitesimal Hubbard attraction can lead to the Cooper instability as long as the density of state is nonzero at the Fermi level, which provides the basis of the SC on the QC. Our numerical results yield that the Cooper pairing always takes place between the two time-reversal states, satisfying the Anderson's theorem. On this theorem, we perform a mean-field (MF) study at both zero and finite temperatures. The MF study also shows that an arbitrarily weak attraction can lead to the pairing order, with the resulted pairing state being well described by the BCS theory and the thermal dynamic behaviors being well consistent with experimental results. The second issue is about the superfluid density on the QC without translational symmetry. It's clarified that although the normal state of the system locates at the critical point of the metal-insulator transition, the pairing state exhibits real SC, carrying finite superfluid density that can be verified by the Meissner effect, consistent with experiment also. These revealed properties of the SC on the Penrose lattice are universal for all QCs.
\end{abstract}


\maketitle

\section{Introduction}
The quasi-crystal (QC) represents a regular type of lattice structure that possesses a certain form of long-range order but lacks translational symmetry \cite{Shechtman1951QCorder, Goldman1993QC}. One famous QC structure is the one-dimensional Fibonacci chain composed of a long ($L$) stick and a short ($S$) one, created by repeating the substitution of $L\rightarrow LS$ and $S\rightarrow L$ \cite{Goldman1993QC, Senechal1996QC}. For two- or three-dimensional QCs, hundreds of materials have been found in metal alloys, especially in aluminum alloys \cite{Goldman1993QC}. These QCs often have an axis with five-, eight-, ten-, or twelve-fold local symmetric axes, which are forbidden in periodic lattices \cite{Shechtman1951QCorder}. Various interesting properties have been revealed about the electron states on the QC, including magnetic order \cite{Wessel2003QC-QAF, Thiem2015QC-QO, Koga2017QC-AFO}, quantum phase transition and criticality \cite{Shaginyan2015QC-QPT, Otsuki2016QC-QCB, Watanabe2016QC-QCB}, strong-correlation behavior \cite{Takemori2015QC-QC, Takemura2015QC-QF, Andrade2015QC-QB}, and topological phases \cite{Kraus2012QC-TS, Huang2018QC-QSHS, Huang2019QC-QSHS, Longhi2019QC-TPT}. Here, we focus on the superconductivity (SC) on the QC \cite{Wong1987EXP, Wagner1988EXP, Deguchi2015EXP, Kamiya2018qSC}, which has caught a lot of interests recently \cite{Sakai2017QC-SC, Hou2018Superfluid, Autti2018Superfluid, Araujo2019QC-SC, Sakai2019pairing, Takemori2020, Nagai2020BdG}.

Recently, definite experimental evidence for the SC was revealed in the Al-Zn-Mg QC with five-folded symmetric axes \cite{Kamiya2018qSC}. The evidence together with those in previous ternary QCs \cite{Wong1987EXP,Wagner1988EXP} and crystalline approximates \cite{Deguchi2015EXP}, has attracted a lot of research interests. Sakai and his coworkers  studied the extended-to-localized crossover of Cooper pairs on the Penrose lattice \cite{Sakai2017QC-SC}. The pairing state for electrons moving in the quasi-periodic potential of the Ammann-Beenker tiling was studied by the Bogoliubov-de Gennes (BdG) approach \cite{Araujo2019QC-SC}, wherein conventional SC consistent with the BCS theory was found. In Ref.~\cite{Nagai2020BdG}, a new numerical skill was developed to treat the BdG equation associated with the SC of the Penrose lattice. However, there are still a few fundamental issues for the SC on the QC that remain to be settled, which are the focus of the present work.

The first issue is the Cooper instability under an infinitesimal attractive interaction. It's well known that on a periodic lattice, a pair of electrons with opposite momenta and spins near the Fermi surface (FS) will be induced by an arbitrarily weak attractive interaction to form a bound state, dubbed as the Cooper pair \cite{Cooper1956}. Such an insight provides a solid basis for the succeeding Bardeen-Cooper-Schrieffer (BCS) theory for the SC \cite{Bardeen1957Theory}. In comparison with a pair of isolate electrons in free space, the presence of a FS is the key ingredient for the Cooper instability. Here in the QC without a FS, will the Cooper instability still be universal for any weak attractive interaction? The second issue is about the superfluid density and the Meissner effect \cite{SchriefferBook}. Although the MF calculations here yield a nonzero pairing gap, it's a problem whether the superconducting phase coherence can survive the disorder-like scattering of the non-periodic lattice. Intuitively, one might wonder how the super current can freely flow through the non-periodic QC lattice, where the momentum is no longer a good quantum number. Actually, there is a basic fact about the transport property on the QC: the normal-state conductivity is critical \cite{Tsunetsugu1991Conductance}, i.e. it decays with the size in power law and converges to zero in the thermal dynamic limit, which means that a macro electronic system on the QC is at the metal-insulator-transition critical point. This knowledge naturally leads to the problem: will the Cooper pairing obtained on the QC lead to real SC with detectable Meissner effect?

In this paper, we study the Cooper instability and SC in the negative-$U$ Hubbard model on the Penrose lattice. Our main results lie in the following three aspects. Firstly, we study a pair of electrons outside a filled Fermi-sea, and investigate their ground state under the Hubbard attraction $U$. Consequently, we analytically prove that an infinitesimal $U$ will lead to Cooper pairing as long as the density of state (DOS) is nonzero at the Fermi level. This result generalizes the Cooper instability from periodic lattices to QCs, providing the SC basis of QCs. Secondly, we perform a mean-field (MF) study for the model at both zero and finite temperatures. Our MF result at the zero temperature is consistent with that of the two-body problem: an infinitesimal attraction can lead to a nonzero pairing order parameter, analytically proved. The MF result at the finite temperature suggests that the thermal dynamic properties of the pairing state can be well described by the BCS theory and are well consistent with the recent experimental results in the  Al-Zn-Mg QC superconductor \cite{Kamiya2018qSC}. Finally, we obtain a nonzero value for the superfluid density of the pairing state, which leads to the real SC with measurable Meissner effect, consistent with the experimental results also \cite{Kamiya2018qSC}.

The remaining part of this work is organized as follows. In Sec.~\ref{model}, we introduce the attractive-$U$ Hubbard model on the Penrose lattice. In Sec.~\ref{ATCI}, we study the two-electron problem outside the Fermi sea to show that an infinitesimal attractive interaction can lead to the Cooper instability. In Sec.~\ref{results}, we provide our MF results at both zero and finite temperatures, to show that the SC of the QC can be well described by the BCS theory. In Sec.~\ref{superfluidity}, the superfluid density is studied, where we show that the pairing state obtained is real SC with finite superfluid density. At last, a brief conclusion is summarized in Sec.~\ref{discussionandconclusion} with some discussions.

\begin{figure}
\centering
\includegraphics[width=0.46\textwidth]{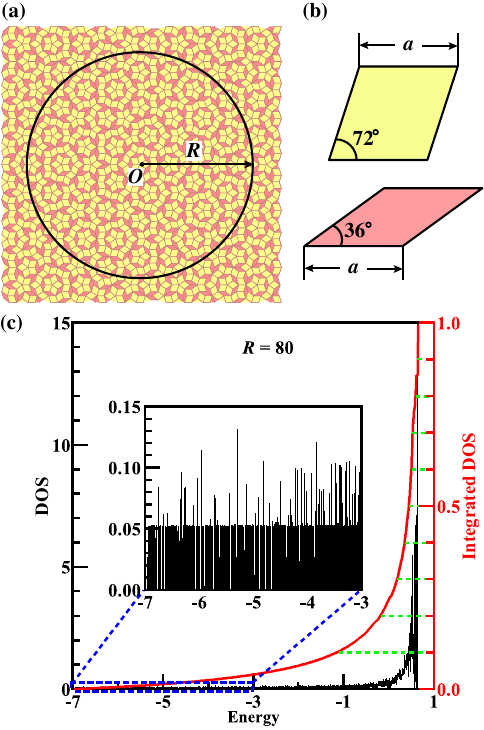}
\caption{(a) The Penrose lattice and (b) its ingredients: the fat and slender rhombis with interior angles of $72^\circ$ and $36^\circ$, respectively. They have the same side length $a$. The original point $O$ in (a) is the center of the five-fold rotational symmetry and $R$ denotes the radius of the considered region. (c) DOS (left axis) and corresponding integrated DOS (right axis). The inset gives the detail of the DOS in the low-filling region.}\label{fig-strucTB}
\end{figure}

\section{model}\label{model}

The Penrose lattice represents a two-dimensional QC, see Fig.~\ref{fig-strucTB}(a), whose original point is the center of the five-fold rotational symmetry. The parameter $R$ gives the radius of the considered region. There are two types of rhombic tilings in the lattice, as shown in Fig.~\ref{fig-strucTB}(b): the fat one with an interior angle of $72^\circ$ and the slender one with an interior angle of $36^\circ$. They have the same side length, $a$. As $R$ increases, the number $N$ of the sites enclosed in the circular region with radius $R$ is roughly proportional to the square of $R$, i.e. $N\sim \left(\sqrt{5}-1\right)\pi R^2/a^2$. In this work, three cases with $R/a=40$, 60, and 80 are adopted to show the size dependence, whose site numbers are $N=6171$, 13926, and 24751 respectively.

Here we consider the following TB model,
\begin{align}
H_{\rm TB}=-\sum_{ij,\sigma}t_{ij}\hat c_{i\sigma}^\dag \hat c_{j\sigma}-\mu_c\sum_{i,\sigma}\hat c_{i\sigma}^\dag \hat c_{i\sigma},
\label{HTB}
\end{align}
where $\hat c_{i\sigma}^\dag$ $\left(\hat c_{i\sigma}\right)$ is the creation (annihilation) operator of the electron with spin $\sigma$ on the $i$-th site and $\mu_c$ is the chemical potential. The hopping integral $t_{ij}$ between the $i$-th and $j$-th sites reads
\begin{align}
t_{ij} = e^{-|\bm r_i-\bm r_j|/a}-\delta_{ij},
\end{align}
which implies a zero on-site energy, i.e. $t_{ii}=0$. Note that the shortest distance between any two sites on this lattice is equal to $\left(\sqrt{5}-1\right)a/2\approx0.618a$, see Fig.~\ref{fig-strucTB}(b). By direct diagonalization, $H_{\rm TB}$ in Eq.~\eqref{HTB} can be rewritten as
\begin{align}
H_{\rm TB}=\sum_{m,\sigma}\tilde\varepsilon_m\hat c_{m\sigma}^\dag \hat c_{m\sigma},
\label{HTBM}
\end{align}
where $\tilde\varepsilon_m\equiv\varepsilon_m-\mu_c$ represents the energy of the state $|m\sigma\rangle$ relative to the chemical potential $\mu_c$. The creation operator of $\hat c_{m\sigma}^\dag $ is defined as
\begin{align}
\hat c_{m\sigma}^\dag=\sum_{i}\xi_{im}\hat c_{i\sigma}^\dag,
\label{cmdag}
\end{align}
where $\xi_{im}\in R$ provides the spatial part of the wave function of the state $|m\sigma\rangle$, satisfying
\begin{equation}\label{xi_normalize}
\sum_{i}\xi_{im}^2=\sum_{m}\xi_{im}^2=1
\end{equation}

As shown in the inset of Fig.~\ref{fig-strucTB}(c), the DOS in the low-filling-fraction region is small and possesses self-similarity, while it sharply increases with the enhancement of the filling fraction. This character is also seen from the integrated DOS, i.e. the red solid line in Fig.~\ref{fig-strucTB}(c). Such a doping-dependence of the DOS suggests that the SC is more favored in the high-filling region.  As examples, we take three chemical potentials of $\mu_c=0.19,\ 0.45,\ {\rm and}\ 0.50$ without loss of generality in the following study. Their filling fractions are $\delta\approx0.3,\ 0.5,\ {\rm and}\ 0.6$ with the corresponding DOSs to be $\varrho\approx0.45, \ 1.40,\ {\rm and}\ 2.53$, respectively.

To study the SC on the Penrose lattice, the following attractive Hubbard interaction is adopted,
\begin{align}
H_{\rm {int}}=-U\sum_i \hat n_{i\uparrow} \hat n_{i\downarrow},
\label{HUI}
\end{align}
where $\hat n_{i\sigma}\equiv\hat c_{i\sigma}^\dag \hat c_{i\sigma}$. This Hamiltonian can be transformed to the eigen-basis representation of $H_{\text{TB}}$ as
\begin{align}
H_{\rm {int}}= -{U\over N} \sum_{mn, m'n'}f_{mn,m'n'}\hat c_{m\uparrow}^\dag \hat c_{n\downarrow}^\dag \hat c_{m'\downarrow} \hat c_{n'\uparrow},
\label{HUM}
\end{align}
where
\begin{align}
f_{mn,m'n'} \equiv N\sum_i \xi_{im}\xi_{in}\xi_{im'}\xi_{in'}= f_{m'n',mn} .
\label{HA}
\end{align}
The total Hamiltonian of the system reads,
\begin{align}
H = H_{\text{TB}}+H_{\rm {int}},
\label{HT}
\end{align}
which sets our start point.

\section{Cooper instability}\label{ATCI}

The Cooper instability \cite{Cooper1956} is the basis of the BCS theory \cite{Bardeen1957Theory} on the periodic lattices. This instability tells about the fate of the two electrons with arbitrarily weak attractive interactions on the background of a filled Fermi sea, that is, their ground state is a bound state with total momenta and spin to be both zero, with an energy lower than zero (relative to the Fermi level) by a finite gap. Such a bound state is called as the Cooper pair. The condensation of the Cooper pairs leads to the SC \cite{Bardeen1957Theory}. In comparison with the case of two electrons in a free space, the presence of a FS as a boundary of the filled Fermi sea is the key ingredient for the Cooper instability on a periodic lattice. However, for the Penrose lattice where the momentum is no longer a good quantum number, the Cooper instability under an infinitesimal attractive interaction is still an issue to be investigated, which is the focus of this section.

Let's consider the two electrons with opposite spins in the background of a filled Fermi sea. The Pauli exclusion principle requires that the single-particle energy of each electron should be higher than the Fermi energy. As a result, the candidate ground-state wave function of this two-electron problem should take the following formula,
\begin{align}
|\Psi_{\rm A}\rangle=\sum_{mn}^{\tilde\varepsilon_{m,n}>0} a_{mn} \hat c_{m\uparrow}^\dag \hat c_{n\downarrow}^\dag |{\rm FS}\rangle,
\label{PsiA}
\end{align}
where $|{\rm FS}\rangle$ represents the filled Fermi-sea state, and the set of real coefficients $\{a_{mn}\}$ satisfy the normalized condition,
\begin{align}
\sum_{mn}^{\tilde\varepsilon_{m,n}>0}a_{mn}^2=1.
\label{amn2}
\end{align}
The sum condition $\tilde\varepsilon_{m,n}>0$ denotes that $\tilde\varepsilon_m$ and $\tilde\varepsilon_n$ are both larger than zero.
The problem now becomes the minimization of the expectation value of the Hamiltonian \eqref{HT} under the trial state in Eq.~\eqref{PsiA}, where the normalized $\{a_{mn}\}$ are the variational parameters.

In the following, we first consider a special case of the wave function \eqref{PsiA} satisfying the condition $a_{mn}=a_{mm}\delta_{mn}$, i.e.,
\begin{align}
|\Psi_{\rm C}\rangle=\sum_m^{\tilde\varepsilon_m>0} a_m \hat c_{m\uparrow}^\dag \hat c_{m\downarrow}^\dag |{\rm FS}\rangle,
\label{PsiC}
\end{align}
with the constraint
\begin{align}
\sum_m^{\tilde\varepsilon_m>0} a_m^2=1.
\label{ConstraintC}
\end{align}
We shall provide an analytical proof that for any weak $U>0$, one can always find a two-electron state described by Eq.~\eqref{PsiC} with the constraint \eqref{ConstraintC} whose energy is below zero by a finite gap in the thermal dynamic limit as long as the DOS at the Fermi level is nonzero, suggesting the formation of a two-electron bound state. In this two-electron bound state, each up-spin single-electron state labeled by $|m{\uparrow}\rangle$ can only be paired with its TR-partner, i.e. the down-spin state labeled by $|m{\downarrow}\rangle$. Such a pairing satisfies the Anderson's theorem \cite{Anderson}. If the minimized energy among this special class of states in Eq.~\eqref{PsiC} is already lower than zero, that among the more general class in Eq.~\eqref{PsiA} should be no higher.

The variational energy, i.e. the expectation value of the Hamiltonian \eqref{HT} in the two-electron trial state, see Eq.~\eqref{PsiC}, can be written as
\begin{align}
E_C=2\sum_m^{\tilde\varepsilon_m>0} a_m^2\tilde\varepsilon_m - {U\over N} \sum_{m,n}^{\tilde\varepsilon_{m,n}>0}f_{mn}a_m a_n,
\label{EC}
\end{align}
with
\begin{align}
f_{mn}\equiv N\sum_i\xi_{im}^2\xi_{in}^2=f_{nm}.
\label{fmn}
\end{align}
Minimizing $E_C$ under the constraint \eqref{ConstraintC} leads to the following self-consistent equation for $\{a_{m}\}$,
\begin{align}\label{self_consistent1}
\frac{\partial}{\partial a_m}\left(2\sum_m^{\tilde\varepsilon_m>0} a_m^2\tilde\varepsilon_m - {U\over N} \sum_{m,n}^{\tilde\varepsilon_{m,n}>0}f_{mn}a_m a_n - \lambda \sum_m^{\tilde\varepsilon_m>0} a_m^2\right)=0,
\end{align}
that is,
\begin{align}\label{self_consistent2}
2a_m\tilde\varepsilon_m - {U\over N} \sum_{n}^{\tilde\varepsilon_n>0}f_{mn}a_n=\lambda a_m.
\end{align}
Note that the Lagrangian multiplier $\lambda$ is just equal to $E_C$ when Eq.~\eqref{self_consistent1} or \eqref{self_consistent2} is satisfied because,
\begin{align}
\lambda = \lambda\sum_m^{\tilde\varepsilon_m>0} a_m^2 =\sum_m^{\tilde\varepsilon_m>0}2a_m^2\tilde\varepsilon_m - {U\over N} \sum_{mn}^{\tilde\varepsilon_{m,n}>0}f_{mn}a_ma_n = E_C.
\end{align}
In the following, we shall prove that for arbitrarily weak $U>0$, there always exists a nonzero solution $\{a_{m}\}$ satisfying Eq.~\eqref{self_consistent2} with finite $\lambda=E_C<0$ in the thermal dynamic limit as long as the DOS at the Fermi level is nonzero. This suggests the formation of a bound state with a finite energy gap, i.e. the Cooper pair. For this purpose, we rewrite Eq.~\eqref{self_consistent2} into the following form,
\begin{align}
A_m = U\sum_n^{\tilde\varepsilon_n>0} F_{mn}^C A_n,
\label{sceqAm}
\end{align}
where
\begin{align}
A_m &\equiv a_m\sqrt{2\tilde\varepsilon_m-\lambda},
\label{Am}\\
F_{mn}^C&\equiv{1\over N}{f_{mn}\over\sqrt{(2\tilde\varepsilon_m-\lambda)(2\tilde\varepsilon_n-\lambda)}}.
\label{Fmn}
\end{align}
Here only the possible candidate states with $\lambda=E_C<0$ are considered. The equation \eqref{sceqAm} takes the form of the eigenvalue problem of the Hermition matrix $F^C$ whose elements are $F_{mn}^C$. The largest eigenvalue of $F^C$ attains $\frac{1}{U}$ and the corresponding eigenvector $A$ determines $\{a_{m}\}$ through Eq.~\eqref{Am}. Below we shall prove that the largest positive eigenvalue of $F^C$ diverges in the limit of $\lambda\to 0^{-}$. Therefore, by properly tuning $\lambda$ to a finite negative value, the largest positive eigenvalue of $F^C$ can certainly attain $\frac{1}{U}$ for any weak $U$, suggesting the formation of the Cooper pair.

Let's consider the following column vector,
\begin{align}
\psi = {1\over \sqrt{Z_\psi}}\left(\cdots, {1\over\sqrt{2\tilde\varepsilon_{m}{-}\lambda}}, \cdots, {1\over\sqrt{2\tilde\varepsilon_{n}{-}\lambda}}, \cdots\right)_{\tilde\varepsilon_{m,n}>0}^T,
\label{psi}
\end{align}
with
\begin{align}
Z_\psi = \sum^{\tilde\varepsilon_{m}>0}_m {1\over 2\tilde\varepsilon_{m}{-}\lambda}.
\label{Zpsi}
\end{align}
Taking $\psi$ as a quantum state $|\psi\rangle$ and $F^C$ as an operator $\hat{F}^C$, let's calculate the expectation value of $\hat{F}^C$ in the state $|\psi\rangle$. The expectation value  $\bar F^C$ is given by
\begin{align}
\bar F^C &\equiv \left\langle\psi\left|\hat{F}^{C}\right|\psi\right\rangle =\psi^{T}F^{C}\psi \nonumber\\&=\frac{1}{NZ_{\psi}}\sum_{mn}^{\tilde\varepsilon_{m,n}>0} \frac{f_{mn}}{(2\tilde\varepsilon_m-\lambda)(2\tilde\varepsilon_n-\lambda)}.
\end{align}
Substituting Eq.~\eqref{fmn} into the above formula, we get
\begin{align}
\bar F^C &={1\over NZ_\psi}\sum_{mn}^{\tilde\varepsilon_{m,n}>0} \frac{N\sum_{i}\xi_{im}^2\xi_{in}^2}{(2\tilde\varepsilon_m-\lambda)(2\tilde\varepsilon_n-\lambda)}\nonumber\\
&={1\over NZ_\psi}N\sum_i\left(\sum_m^{\tilde\varepsilon_m>0} {\xi_{im}^2\over 2\tilde\varepsilon_m-\lambda}\right)\cdot\left(\sum_n^{\tilde\varepsilon_n>0} {\xi_{in}^2\over 2\tilde\varepsilon_n-\lambda}\right)\nonumber\\
&={1\over NZ_\psi}\sum_i 1^2\times\sum_i\left(\sum_m^{\tilde\varepsilon_m>0} {\xi_{im}^2\over 2\tilde\varepsilon_m-\lambda}\right)^2\nonumber\\
&\geqslant{1\over NZ_\psi}\left(\sum_i\sum_m^{\tilde\varepsilon_m>0} {\xi_{im}^2\over 2\tilde\varepsilon_m-\lambda}\right)^2={1\over NZ_\psi}\left(\sum_m^{\tilde\varepsilon_m>0}{\sum_i\xi_{im}^2\over 2\tilde\varepsilon_m-\lambda}\right)^2,
\end{align}
where the Cauchy's inequality is used. Substituting Eq.~\eqref{xi_normalize} and \eqref{Zpsi} into the above formula, we have
\begin{align}
\bar F^C &\geqslant{1\over NZ_\psi}\left(\sum_m^{\tilde\varepsilon_m>0}{1\over 2\tilde\varepsilon_m-\lambda}\right)^2={1\over N}\sum_m^{\tilde\varepsilon_m>0} {1\over 2\tilde\varepsilon_{m}{-}\lambda}\nonumber\\&\xrightarrow{N\to\infty}\int_0^{\infty}{\varrho(\tilde\varepsilon)\over 2\tilde\varepsilon-\lambda}d\tilde\varepsilon.
\end{align}
Here $\varrho$ represents the DOS. In the limit of $\lambda\to 0^{-}$, we have
\begin{align}\label{proof}
\bar F^C&\geqslant\varrho(0)\int_0^{0^+}{1\over 2\tilde\varepsilon}d\tilde\varepsilon \rightarrow +\infty,
\end{align}
as long as the DOS at the Fermi level, $\varrho(0)$, is nonzero.

Since the expectation value $\bar {F}^C$ of the Hermitian operator $\hat {F}^C$ in the constructed state $|\psi\rangle$ diverges in the limit of $\lambda\to 0^{-}$, the largest positive eigenvalue of $\hat F^C$, which should be no less than $\bar {F}^C$, must also diverge in that limit. Therefore, for however weak $U$, there always exists a finite negative $\lambda$ dictating that the largest eigenvalue of $\hat F^C$ attains $\frac{1}{U}$, satisfying Eq.~\eqref{sceqAm}. The parameter $\lambda=E_C<0$ is the minimized energy among the special variational class of states described by Eq.~\eqref{PsiC}. Thus, the minimized energy among the more general variational class of states described by Eq.~\eqref{PsiA}, which should be no higher than $\lambda$, is also negative. Note that the single-particle state $|m\sigma\rangle$ on the Penrose lattice is critical \cite{Tsunetsugu1991Wavefunction}, instead of localized. This means that to make the energy of the state \eqref{PsiA} negative, it should be a two-electron bound state, i.e. the Cooper pair. To this point, we have proved that an infinitesimal attractive Hubbard interaction can lead to the Cooper instability on the QC, once the DOS at the Fermi level is finite.

Note that the DOS of the electron states on a QC exhibits a fractal character \cite{Tsunetsugu1991Wavefunction}. Concretely, the DOS curve is singular smooth: on the one hand, it contains a singular part manifested as sharp infinite-height peaks here and there; on the other hand, the integrated DOS curve is smooth, suggesting that the energy-level points for the sharp DOS peaks form no measure. Such a fractal character of the DOS leads to pseudogaps and sharp peaks here and there in the DOS curve. In addition, the DOS curve can contain a smooth part superposed on the singular part \cite{Tsunetsugu1991Wavefunction}, which leads to a finite DOS background. Therefore, the pseudogaps at most places in the DOS curve are not real gaps and the DOS there is nonzero. Note that due to the singular part in the DOS curve, the DOS is not mathematically vigorously defined. However, on the above proof, we only require that the averaged DOS in an infinitesimal energy shell near the Fermi level is larger than zero, as embodied in Eq. (\ref{proof}), which is satisfied at most places in the DOS spectrum.  Of course, under such a singular energy dependence of the DOS spectrum, the properties of the SC, including the $T_c$, the pairing gap, the superfluid density, etc, will exhibit a very sensitive dependence on the filling fraction. However, in real materials, the presence of a weak randomness can largely smear out the singular part of the DOS spectrum \cite{Yamamoto1995}, leading to the much smoother filling-fraction dependence of the superconducting properties of the system. In the experiment of the Al-Zn-Mg QC \cite{Kamiya2018qSC}, the linear-dependence of the specific heat with temperature at low temperature suggests a finite DOS at the Fermi level, which thus satisfies the condition required here for the Cooper instability.

To obtain the optimized Cooper-pair wave function, we consider the general variational state in Eq.~\eqref{PsiA}. The expectation value of the Hamiltonian \eqref{HT} in this state reads,
\begin{align}
E_{\rm A}=\sum_{mn}^{\tilde\varepsilon_{m,n}>0} a_{mn}^2(\tilde\varepsilon_m + \tilde\varepsilon_n)- {U\over N} \sum_{mn,m'n'}^{\tilde\varepsilon_{m,n,m',n'}>0}f_{mn,m'n'}a_{mn} a_{m'n'}.
\label{EA}
\end{align}
The sum condition $\tilde\varepsilon_{m,n,m',n'}>0$ denotes that $\tilde\varepsilon_m$, $\tilde\varepsilon_n$, $\tilde\varepsilon_{m'}$, and $\tilde\varepsilon_{n'}$ are all larger than zero.
Minimizing $E_A$ under the constraint \eqref{amn2},
\begin{align}\label{self_consistentmn}
\frac{\partial}{\partial a_{mn}}\left(\sum_{mn}^{\tilde\varepsilon_{m,n}>0} a_{mn}^2(\tilde\varepsilon_m + \tilde\varepsilon_n)- {U\over N} \sum_{mn,m'n'}^{\tilde\varepsilon_{m,n,m',n'}>0}f_{mn,m'n'}a_{mn} a_{m'n'}\right.\nonumber\\
\left. - \lambda \sum_{mn}^{\tilde\varepsilon_{m,n}>0} a_{mn}^2\right)=0,
\end{align}
leads to the following self-consistent equation for the set of $\{a_{mn}\}$,
\begin{align}\label{self_consistent_mn1}
\left(\tilde\varepsilon_m + \tilde\varepsilon_n\right)a_{mn} - {U\over N} \sum_{m',n'}^{\tilde\varepsilon_{m', n'}>0}f_{mn, m'n'}a_{m'n'}=\lambda a_{mn}.
\end{align}
Here, again the Lagrange multiplier $\lambda$ is equal to $E_A$ when the self-consistent equation is satisfied. The equation \eqref{self_consistent_mn1} takes the form of the eigenvalue problem of the Hermitian matrix $F_{mn,m'n'}^{A}$,
\begin{align}\label{self_consistent_mn2}
\sum_{m'n'}^{\tilde\varepsilon_{m'n'}>0}F_{mn,m'n'}^{A}a_{m'n'}=\lambda a_{mn}
\end{align}
with
\begin{align}\label{F_mnm'n'}
F_{mn,m'n'}^{A}=\left(\tilde\varepsilon_m + \tilde\varepsilon_n\right)\delta_{mn,m'n'} - {U\over N}f_{mn, m'n'},
\end{align}
which can be solved numerically.

\begin{figure}
\centering
\includegraphics[width=0.46\textwidth]{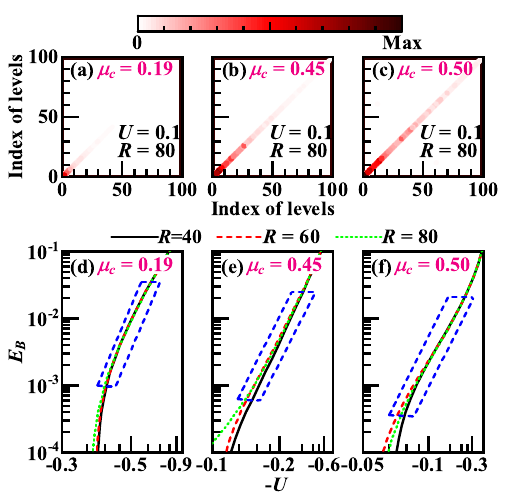}
\caption{(a-c) The distribution of $a_{mn}^2$ in the $m$-$n$ plane for the one hundred energy levels marked by $m$ and $n$ above the Fermi level. The $x$- and $y$- axes represent for $m$ and $n$ and the color represents for $a_{mn}^2$. For (a), (b), and (c), $U=0.1$ and $R=80$, but with three different doping levels whose $\mu_c=0.19,0.45$ and $0.5$, respectively.  (d-f) The binding energy $E_B$ as function of the attractive interaction strength $U$ for the three different doping levels and three different sizes ($R=40$, 60, and 80). The vertical axis adopts the logarithmic coordinate, while the horizontal one adopts reciprocal coordinates for $-U$.}
\label{fig-cooperpair}
\end{figure}

Considering one hundred states above the Fermi level, the numerical solution of Eq.~\eqref{self_consistent_mn2} is obtained. For each doping and $U$, the ground state of this two-body system is a bound-state, whose wave function is plotted in Figs.~\ref{fig-cooperpair}(a-c) for the three doping levels with $\mu_c=0.19,0.45$ and $0.5$, where the $x$ and $y$ axes represent $m$ and $n$ and the color represents $a_{mn}^2$. From Figs.~\ref{fig-cooperpair}(a)-\ref{fig-cooperpair}(c), it's clear that for each $m$, we have $|a_{mm}|\gg |a_{mn}|_{n\ne m}$. Such a solution makes the general wave function \eqref{PsiA} decay to the special one \eqref{PsiC} that satisfies the Anderson's theorem. In a small energy scale, the fractal structure of the DOS, together with some finite-size effect, is responsible for the non-monotonic dependence of the pairing amplitudes on the index of levels or energy (levels are arranged in the ascending order of energy). When the energy scale is large enough, the pairing magnitudes decay with the index of levels in a whole, see Figs.~\ref{fig-cooperpair}(a)-\ref{fig-cooperpair}(c), suggesting that the Cooper pairing is mainly contributed by the states near the Fermi level. This is also reflected in the increasing of the length of the red line-shape region from Fig.~\ref{fig-cooperpair}(a) to Fig.~\ref{fig-cooperpair}(c), because the energy range that covers the considered 100 states decreases for increasing the DOS. These results suggest that the Cooper pairing only takes place between the two TR states near the Fermi level, satisfying the Anderson's theorem \cite{Anderson}.

We further investigate the binding energy $E_B\equiv |E_A|=|E_C|$ of the Cooper pair, i.e., the bound-state gap, which reflects the strength of the pairing. Figures \ref{fig-cooperpair}(d)-\ref{fig-cooperpair}(f) show $E_B$ as a function of $U$, where the vertical axis adopts the logarithmic coordinate for $E_B$ while the horizontal one adopts the reciprocal coordinate for $-U$. From Figs.~\ref{fig-cooperpair}(d)-\ref{fig-cooperpair}(f), it's clear that $E_B$ reasonably enhances with the enhancement of $U$. We focus on the regimes framed in the blue parallelogram in Figs.~\ref{fig-cooperpair}(d)-\ref{fig-cooperpair}(f) wherein, on the one hand, the binding energy is much larger than the finite-size level spacing so that the thermal-dynamic-limit behavior is shown; on the other hand, the Hubbard-$U$ is not so strong. Consequently, in this regime, $\ln(E_B)$ linearly depends on $-\frac{1}{U}$, suggesting
\begin{align}
E_B \propto \exp\left[-{1\over \alpha U}\right].
\label{EB1}
\end{align}
This result is consistent with the BCS theory for the periodic lattice \cite{Cooper1956, Bardeen1957Theory}. In the latter case, we further have $\alpha\propto\varrho(0)$. This relation is qualitatively consistent with our results here, because the slope of the linear-dependence relation between  $\ln(E_B)$ and $-\frac{1}{U}$ shown in Fig.~\ref{fig-cooperpair}(d)-\ref{fig-cooperpair}(f) decreases with the doping and hence $\varrho(0)$. However, due to the finite size adopted in our calculations, we cannot quantitatively check this relation.

\section{Mean-field results}\label{results}

On the above, we prove that an infinitesimal Hubbard interaction on the Penrose lattice would lead to the Cooper instability, which provides the basis of the SC in the system.  In this section, we shall perform a MF study for the system, at both the zero and finite temperatures. Our zero-temperature MF study further confirms the above result: an infinitesimal Hubbard attraction would lead to the pairing order. Our finite-temperature results reveal that the thermal dynamic behavior of the superconducting state on the QC can be well described by the BCS theory.

\subsection{Zero-temperature Results}

The last section shows that the Cooper pairing formed by the two electrons outside the Fermi sea obeys the Anderson's theorem, i.e., the single particle state $|m{\uparrow}\rangle$ can only be paired with its TR-partner $|m{\downarrow}\rangle$. Accordingly, the MF state with order parameter $\left\langle c_{m\uparrow}c_{m\downarrow}\right\rangle$ naturally emerges as the result of As a result of the condensation of the Cooper pairs. The MF decomposition of the Hamiltonian \eqref{HT} in this channel leads to the following BdG Hamiltonian,
\begin{equation}
H_{\rm BdG}=\sum_{m\sigma}\tilde\varepsilon_m \hat c_{m\sigma}^\dag \hat c_{m\sigma}-\sum_m\left(\Delta_m\hat c_{m\uparrow}^\dag \hat c_{m\downarrow}^\dag + {\rm h.c.}\right)+{\rm const.},\label{HM}
\end{equation}
where the pairing order parameters of $\{\Delta_m\}$ are defined as \begin{equation}
\Delta_m={U\over N}\sum_{n}f_{mn}\left\langle\hat c_{n\downarrow}\hat c_{n\uparrow}\right\rangle.
\label{Deltam}
\end{equation}
Using the Bogoliubov transformation,
\begin{subequations}
\begin{align}
\hat c_{m\uparrow}&=u_m\hat\gamma_{m\uparrow} + v_m\hat\gamma_{m\downarrow}^\dag,\\
\hat c_{m\downarrow}&=u_m\hat\gamma_{m\downarrow} - v_m\hat\gamma_{m\uparrow}^\dag,
\end{align}\label{BogTrans}
\end{subequations}
the BdG Hamiltonian can be diagonalized with the Bogoliubov pairing coherence factors,
\begin{align}
u_m=\sqrt{{1\over2}\left(1+{\tilde\varepsilon_m\over E_m}\right)}, \quad v_m={\rm sgn}(\Delta_m)\sqrt{{1\over2}\left(1-{\tilde\varepsilon_m\over E_m}\right)},
\label{uvm}
\end{align}
with
\begin{align}\label{dispersion}
E_m=\sqrt{\tilde\varepsilon_m^2+\Delta_m^2}.
\end{align}
Substituting Eq.~\eqref{BogTrans} into Eq.~\eqref{Deltam}, we have
\begin{align}
\Delta_m = {U\over N}\sum_n f_{mn}u_nv_n\left(1-\langle\hat\gamma_{m\uparrow}^\dag\hat\gamma_{m\uparrow}\rangle -\langle\hat\gamma_{m\downarrow}^\dag\hat\gamma_{m\downarrow}\rangle\right).
\label{sceq0}
\end{align}
At the zero temperature, the above equation turns into the following self-consistent one for $\{\Delta_m\}$,
\begin{align}
\Delta_m = {U\over N}\sum_n f_{mn}{\Delta_n\over 2\sqrt{\tilde\varepsilon_n^2+|\Delta_n|^2}}.
\label{sceqDeltam}
\end{align}

With the same approach adopted in the last section, we can prove that an arbitrarily weak $U$ can lead to finite values of $\{\Delta_m\}$ as long as the DOS at the Fermi level is nonzero. To show this, we first transform Eq.~\eqref{sceqDeltam} into
\begin{align}
\tilde\Delta_m = U\sum_n F_{mn}^M\tilde\Delta_n,
\label{sceqDeltam_symmetric}
\end{align}
with
\begin{align}
\tilde\Delta_m &= {\Delta_m\over\left(\tilde\varepsilon_m^2+|\Delta_m|^2\right)^{1\over 4}},\\
F_{mn}^M &= {f_{mn}\over 2N\left[\left(\tilde\varepsilon_m^2+|\Delta_m|^2\right)\left(\tilde\varepsilon_n^2+|\Delta_n|^2\right)\right]^{1\over4}}.
\label{F_mn}
\end{align}
Again, the equation \eqref{sceqDeltam_symmetric} takes the form of the eigenvalue problem of the Hermitian matrix $F^M$, wherein the largest eigenvalue of $F^M$ attains $\frac{1}{U}$. Note that for $U\to 0$, we have $\Delta_m\to 0$. In this limit, the matrix $F^M$ is just equal to the $F_{mn}^C$ defined in Eq.~\eqref{Fmn} in the limit of $\lambda\to 0^{-}$, whose largest eigenvalue has been proved to diverge in the last section. Hence, for any weak $U$, we can always find a group of weak but finite $\{\Delta_m\}$ so that the largest eigenvalue of $F_{mn}^M$ attains $\frac{1}{U}$, satisfying Eq.~\eqref{sceqDeltam_symmetric}. Therefore, we have proved here that an infinitesimal Hubbard-attraction will lead to the pairing order on the Penrose lattice as long as the DOS at the Fermi level is nonzero.

\begin{figure}[htbp]
\centering
\includegraphics[width=0.46\textwidth]{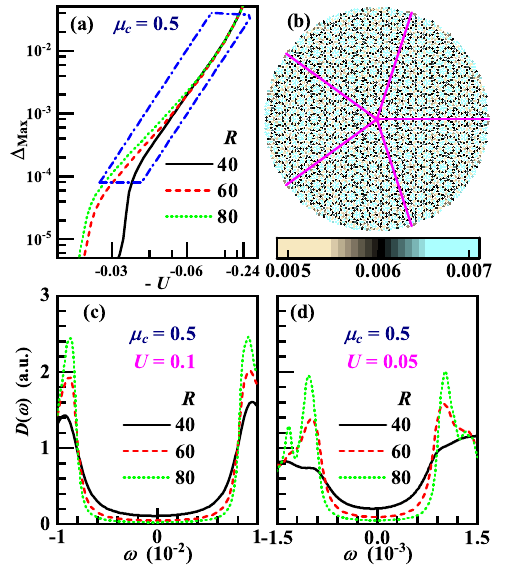}
\caption{(a) The maximum pairing gap $\Delta_{\rm Max}$ as a function of $U$. The horizontal and vertical axes are reciprocal for $-U$ and logarithmic for $\Delta_{\rm Max}$, respectively. (b) The real-space distribution of the pairing gap $\Delta_{i}$. (c) and (d) The tunneling spectra $D(\omega)$ in the superconducting ground state for two different $U$. All the adopted parameters  are shown in each panel.}
\label{fig-scgapZT}
\end{figure}

For a general $U>0$, the self-consistent gap equation \eqref{sceqDeltam} or equally \eqref{sceqDeltam_symmetric} can be solved numerically by iterative method. Figure~\ref{fig-scgapZT}(a) shows the maximum SC gap among $\{\Delta_m\}$, i.e., $\Delta_{\rm Max}$, as a function of $U$ at zero temperature for three different system sizes. Similarly with the results for the two-electron problem provided in Figs.~\ref{fig-cooperpair}(d)-\ref{fig-cooperpair}(f), the framed regime in Fig.~\ref{fig-scgapZT}(a) suggests that for weak $U$ in the thermal dynamic limit, we have
\begin{align}
\Delta_{\rm Max} \propto \exp\left[-{1\over\alpha U}\right],
\label{DeltaMin}
\end{align}
consistent with the BCS theory \cite{Bardeen1957Theory}. Physically, we should have $\alpha\propto \varrho(0)$, which could be tested by the larger lattices. Since $\Delta_{m}$ in Eq.~\eqref{Deltam} is a sum of the $\left\langle\hat c_{n\downarrow}\hat c_{n\uparrow}\right\rangle$ for all levels, rather than just $\left\langle\hat c_{m\downarrow}\hat c_{m\uparrow}\right\rangle$, the maximum gap $\Delta_{\rm Max}$ and the minimum one $\Delta_{\rm Min}$ are in the same order of magnitude, see the following Figs.~\ref{fig-scgapFT}(a) and \ref{fig-scgapFT}(b).  In Fig.~\ref{fig-scgapZT}(b), the real-space distribution of the pairing gap function $\Delta_i$ is shown, which reads as
\begin{equation}
\Delta_i=\sum_m\Delta_m\xi_{im}^2.
\end{equation}
The five-folded symmetric pattern illustrated in Fig.~\ref{fig-scgapZT}(b) is consistent with the $s$-wave pairing symmetry.

The SC gap can be measured by the tunneling spectrum, provided as
\begin{align}
D(\omega) \equiv -{\rm Im}\sum_m\left({u_m^2\over \omega-E_m+i0^+}+{v_m^2\over \omega+E_m+i0^+}\right).
\label{Domega}
\end{align}
The $D(\omega)$ for $U=0.1$ and $U=0.05$ are plotted in Figs.~\ref{fig-scgapZT}(c) and \ref{fig-scgapZT}(d), respectively. The clear U-shape curves reflect the full-gap character of the pairing state.

\begin{figure}[htbp]
\centering
\includegraphics[width=0.46\textwidth]{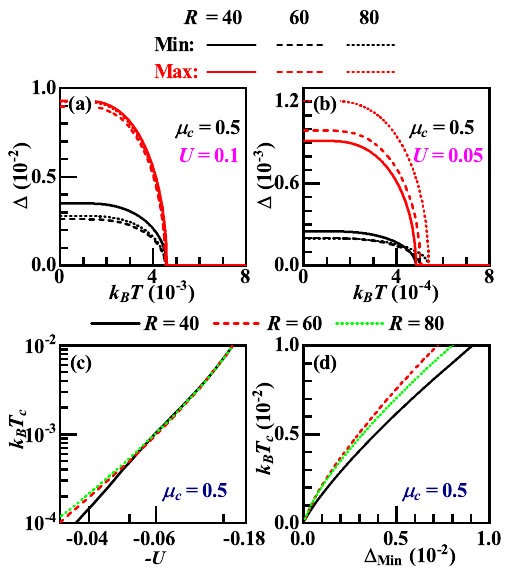}
\caption{The maximum gap $\Delta_{\rm Max}$ and the minimum one $\Delta_{\rm Min}$ as function of $T$ for three different lattice sizes with $R=40$, 60, and 80 for (a) $U=0.1$ and (b) $U=0.05$. The $U$-dependence of $k_BT_c$ (c) and the relation between $\Delta_{\rm Min}$ and $k_BT_c$ (d). All the used parameters are shown in each panel.}
\label{fig-scgapFT}
\end{figure}

\subsection{Finite temperature}
At the finite temperature $T$, substituting
\begin{align}
\left\langle\hat\gamma_{m\uparrow}^\dag\hat\gamma_{m\uparrow}\right\rangle = \left\langle\hat\gamma_{m\downarrow}^\dag\hat\gamma_{m\downarrow}\right\rangle = {1\over e^{-E_m/k_BT} + 1}
\end{align}
into Eq.~\eqref{sceq0},we obtain the self-consistent equations for $\{\Delta_m\}$ as follow,
\begin{align}
\Delta_m &= {U\over N}\sum_n f_{mn}{\Delta_n\over 2\sqrt{\tilde\varepsilon_n^2+|\Delta_n|^2}}\times \tanh\left({1\over2} {E_n\over k_BT}\right).
\label{sceqDeltamFT}
\end{align}
Here $k_B$ is the Boltzmann constant. This set of equations are solved numerically by iteration approach. The maximum gap $\Delta_{\rm Max}$ and the minimum one $\Delta_{\rm Min}$ as function of $T$ for three different lattice sizes with $R=40$, 60, and 80 are shown in Figs.~\ref{fig-scgapFT}(a) and \ref{fig-scgapFT}(b) for $U=0.1$ and $0.05$, respectively. Obviously, both $\Delta_{\rm Max}$ and $\Delta_{\rm Min}$ decrease with $T$ until at a superconducting critical temperature $T_c$, at which both drop to zero. Note that for the case of $U=0.05$ shown in Fig.~\ref{fig-scgapFT}(b) the $T_c$ exhibits an obvious size-dependent, as the small pairing gap in this case is not far from the finite-size level spacing. The temperature-dependence of $\Delta_{\rm Max}$ and $\Delta_{\rm Min}$ near $T=0$ and $T=T_c$ is consistent with the BCS theory for periodic lattices \cite{Bardeen1957Theory}. Particularly, our detailed analysis suggests that both $\Delta_{\rm Max}$ and $\Delta_{\rm Min}$ scale with $\left(T_c-T\right)^{\frac{1}{2}}$ for $T$ slightly lower than $T_c$. Such a temperature dependent behavior of the pairing gap will lead to an upper critical field $h\propto \left(T_c-T\right)$, being well consistent with the experiment of Al-Zn-Mg superconductor \cite{Kamiya2018qSC}.

The $U$-dependence of $k_BT_c$ is plotted in Fig.~\ref{fig-scgapFT}(c), which also satisfies the similar relation as Eqs.~\eqref{EB1} and \eqref{DeltaMin},
\begin{align}
k_BT_c \propto \exp\left[-{1\over\alpha U}\right].
\label{kBTc}
\end{align}
Varying $U$, the relation between $\Delta_{\rm Min}$ and $k_BT_c$ is shown in Fig.~\ref{fig-scgapFT}(d), from which we find that $k_BT_c\propto\Delta_{\rm Min}$ for weak $U$. The situation is similar for $\Delta_{\rm Max}$. All these temperature-dependent behaviors obtained here are well consistent with the BCS theory \cite{Bardeen1957Theory}.

%%%%%
\begin{figure}[htbp]
\centering
\includegraphics[width=0.46\textwidth]{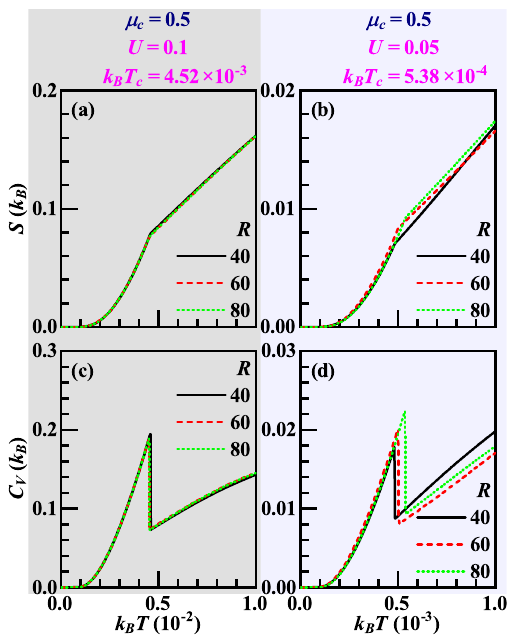}
\caption{The temperature dependence of the entropy (a, b) and specific heat (c, d) for $U=0.1$ (left column) and 0.05 (right column). Three different lattices with sizes $R=40,60$ and $80$ are studied, and other parameters are shown on top.}
\label{fig-SCV}
\end{figure}

To study the thermal dynamic property of the system, especially that of the superconducting phase transition, we have calculated the entropy $S$ and the specific heat $C_V$, which are formulated as \cite{Bardeen1957Theory}
\begin{align}
S &= -k_B{2\over N}\sum_m \left[(1-f_m)\ln(1-f_m)+f_m\ln f_m\right],
\label{entropy}\\
C_V &= (k_BT){\partial \over \partial (k_BT)}S,
\label{entropy}
\end{align}
where $f_m=\left(1+e^{E_m/k_BT}\right)^{-1}$. The temperature-dependences of $S$ and $C_V$ for $U=0.1$ and $0.05$ for the three different lattice sizes are shown in Figs.~\ref{fig-SCV}(a)-\ref{fig-SCV}(d). From Figs.~\ref{fig-SCV}(a) and \ref{fig-SCV}(b), the entropy is continuous at $T_c$, while its first-order derivative is discontinuous, which leads to a jump for the corresponding specific heat shown in Figs.~\ref{fig-SCV}(c) and \ref{fig-SCV}(d). Such a behavior of the temperature-dependence of the specific heat is well consistent with the experiment on the Al-Zn-Mg QC superconductor \cite{Kamiya2018qSC}. Figure \ref{fig-SCV} suggests that the SC transition on the Penrose lattice here is a second-order phase transition, consistent with the BCS theory for periodic lattices \cite{Bardeen1957Theory}.

\begin{figure}
\centering
\includegraphics[width=0.46\textwidth]{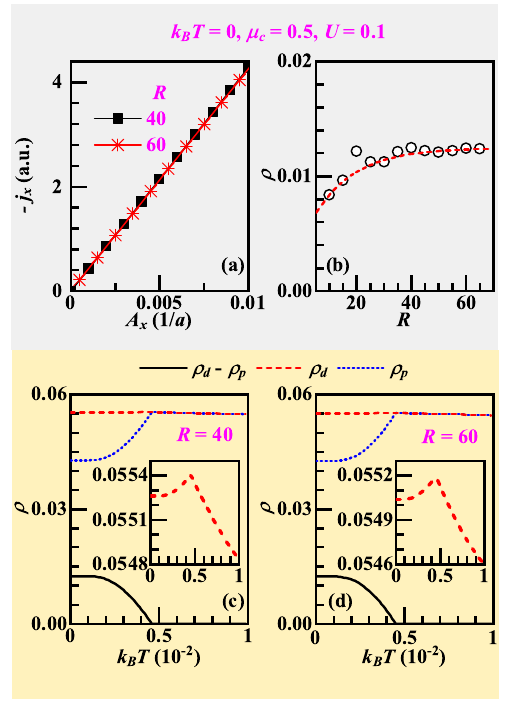}
\caption{Superfluid density of the pairing state. (a) Expectation value of the current operator as function of the vector potential for the pairing state. (b) The size-dependence of the superfluid density. In (b), the scatter dots represent the numerical results, while the dashed lines are for guidance. (c) and (d) Temperature dependences of $\rho_d$, $\rho_p$ and their difference $\rho_d-\rho_p$ for two different lattices with sizes $R = 40$ and $R = 60$, respectively. For easy observation, the temperature dependence of $\rho_d$ is enlarged near $k_BT_c$ and replotted in the insets of (c) and (d).}
\label{fig-sc}
\end{figure}

\section{superfluid density}\label{superfluidity}

The MF calculations in the last section yield a nonzero pairing gap. However, one might wonder whether real SC with measurable Meissner effect can be detected. Intuitively, it needs to be understood how the super current can freely flow through the non-periodic QC lattice where the momentum is no longer a good quantum number. It's known that the electron transport property on the QC is critical \cite{Tsunetsugu1991Conductance}: the normal-state conductivity of the Penrose lattice decays with the size in a power law and converges to zero in the thermal dynamic limit, which means that a macro electronic system on the QC is at the metal-insulator-transition critical point. Therefore, it needs to be clarified whether such a critical state can be driven to real SC by attractive electron-electron interactions. To settle this issue, we should study the superfluid density, whose nonzero value suggests real SC with measurable Meissner effect.

The experimental identification of the SC is the Meissner effect. Physically, the Meissner effect is brought about by the combination of the universal Maxwell equation and the London equation for superconductors \cite{Bardeen1957Theory, SchriefferBook}, i.e. $\left\langle\hat{\bm {j}}\right\rangle=-\rho \bm{A}$. Here $\bm{A}$ represents the weak smooth vector potential imposed on the system and $\left\langle\hat{\bm {j}}\right\rangle$ represents the expectation value of the current operator $\hat{\bm {j}}$ as a response of $\bm{A}$.  The nonzero $\rho$, called as the superfluid density, will lead to real SC with measurable Meissner effect. Theoretically, the superfluid density reflects the phase rigidity of SC. Assuming a weak uniform $\bm{A}$, the site-averaged current operator $\hat{\bm {j}}$ is defined as
\begin{align}
\hat {\bm {j}} \equiv -\frac{1}{N}{\partial \over \partial \bm{A}}H_{\rm TB}(\bm{A}),
\label{JJ}
\end{align}
where $H_{\rm TB}(\bm{A})$ has the form of
\begin{align}
H_{\rm TB}(\bm A) = & -\sum_{ij,\sigma}t_{ij}e^{i\int_{\bm{r}_i}^{\bm{r}_j}\bm A\cdot d\bm l} \hat c_{i\sigma}^\dag \hat c_{j\sigma}.
\label{HA}
\end{align}
The vector potential $\bm A$ influeces the system through revising the hopping integrals of the tight-binding Hamiltonian.
The site-averaged current operator $\hat{\bm {j}}$ can be separated into
\begin{equation}\label{j_total}
\bm {\hat j}  = - \left(\bm {\hat j}^d-\bm {\hat j}^p\right)
\end{equation}
with the site-averaged  paramagnetic current operator and diamagnetic one defined as follow,
\begin{align}
\bm {\hat j}^p&=\frac{i}{2N}\sum_{ij,\sigma}t_{ij}\bm{r}_{ij}c_{i\sigma}^{\dagger}c_{j\sigma}+{\rm h.c.},
\label{j_p}\\
\bm {\hat j}^d&=\frac{1}{2N}\sum_{ij,\sigma}t_{ij}\bm{r}_{ij}(\bm{r}_{ij}\cdot\bm{A})c_{i\sigma}^{\dagger}c_{j\sigma}+{\rm h.c.}.
\label{j_d}
\end{align}
Here $\bm{r}_{ij}\equiv\bm{r}_j-\bm{r}_i$ denotes the vector pointing from the site $i$ to the site $j$. In a general possibly-anisotropic 2D system, the superfluid density $\varrho$ should be a $2\times2$ tensor. However, due to the $D_5$-point group, it can be proved that $\rho$ is simply a number \cite{Cao2020QCSC}. Therefore, we are allowed to orientate the vector potential $\bm A$ along the $x$ axis and calculate the $x$-component of the corresponding current $j_x\equiv\langle\hat{j}_x\rangle$.

The procedure to calculate the expectation value $j_x$ of the current operator $\hat{j}_x$ is as follow. Firstly, we replace the free-electron term in the BCS-MF Hamiltonian (\ref{HM}) by the $H_{\rm TB}(\bm A)$ in Eq.~\eqref{HA} under a small $A_x$ and keep the pairing term in Eq.~(\ref{HM}) unchanged. Secondly, we numerically re-diagonalize the revised Bogoliubov-de-Genes MF Hamiltonian to obtain the revised eigenstates and revised Bogoliubov-quasi-particle eigen energies. Finally, we calculate the thermal expectation value $j_x$ of the current operator $\hat{j}_x$ at a given temperature, including those of the diamagnetic and paramagnetic parts provided by Eqs.~\eqref{j_p} and \eqref{j_d}, respectively. The difference between the expectation values of the two parts gives the final value $j_x$.

In Fig.~\ref{fig-sc}(a), the responding $-j_x$ toward an imposed weak uniform $\bm A=A_x\bm{e}_x$ is shown for two lattices with sizes $R=40$ and 60 under the open boundary condition, where a linear-response relation is obtained with positive slope, suggesting $\rho>0$. In Fig.~\ref{fig-sc}(b), the lattice-size dependence of $\rho$ is shown, which suggests that $\rho$ is saturated to a nonzero value in the thermal dynamic limit. Such a nonzero superfluid density leads to real SC with finite phase rigidity that can be detected by the Meissner effect. Therefore, real SC can indeed emerge in the QC system, although the normal state in the QC locates at a metal-insulator-transition critical point. This result is consistent with the experiment of the Al-Zn-Mg superconductor, where the suppress of the SC by exerted magnetic field just reflects the Meissner effect \cite{Kamiya2018qSC}.

The temperature dependence of $\rho_p$, $\rho_d$ and $\rho$ are shown in Figs.~\ref{fig-sc}(c) and \ref{fig-sc}(d) for $R = 40$ and $R = 60$, respectively. Both the diamagnetic and paramagnetic superfluid densities are nonzero at a general temperature below $T_c$ and they are unequal. However, when the temperature is above $T_c$, the pairing gap closes and the obtained diamagnetic and paramagnetic superfluid densities exactly cancel each other to give zero total superfluid density in the normal state.
It is shown that $\rho_d$ doesn't obviously change with temperature, although there is a weak cusp (see the insets) at $T_c$ . The $\rho_p$ instead is obviously enhanced by temperature. Physically, the enhancement of $\rho_p$ originates from the extra consumption of the superfluid density caused by the Bogoliubov quasi-particle excitations which are largely enhanced by the temperature. As a result, the total superfluid density $\rho=\rho_d-\rho_p$ is suppressed by the enhancement of temperature and vanishes if $T>T_c$, where the SC vanishes, too.

\section{Discussions and Conclusion}\label{discussionandconclusion}

The pairing symmetry of the system obtained here is singlet on-site s-wave. Such a pairing symmetry is determined by the on-site attractive character of the interaction part of the negative-$U$ Hubbard model. For a general interaction form, more pairing symmetries are possible, including the singlet or triplet $s$-wave, $(p_x,p_y)$-wave, $(d_{x^2-y^2},d_{xy})$-wave and $h$-wave pairing symmetries for the Penrose lattice with $D_5$- point group. The scheme for the classification of the pairing symmetries based on the group theory, and the properties of the unconventional pairing states, including the topological pairing states, have been studied in Ref.~\cite{Cao2020QCSC}.

It's important to note that the satisfaction of the Anderson's theorem is crucial for the BCS-like behaviors of the on-site pairing state obtained here. For more general pairing states on the QC, such as those driven by repulsive interactions via the Kohn-Luttinger's mechanism \cite{Cao2020QCSC}, the Anderson's theorem might be broken. As a result, each single-particle state with the state index $m$ can be paired with any other state with index $n$. In such a pairing state, the energy dispersion relation of the Bogoliubov quasi-particles can no longer be described by Eq.~\eqref{dispersion}. Generally, no analytical formula is available for the quasi-particle energy. What's more, the Bogoliubov quasi-particles might in general host a gapless spectrum, just like the one in the normal state. The behavior of such a gapless pairing state can be very different from the one presented here. We leave such a topic for future study.

In conclusion, our systematic study on the attractive Penrose Hubbard model has settled several fundamental issues about the SC on the QC. The first issue is about the Cooper instability for infinitesimal attractions on the QC. We provide a vigorous proof that an infinitesimal Hubbard attraction can lead to the Cooper pairing between the  TR partners, satisfying the Anderson's theorem. This result provides basis for the SC on the QC. Our MF results on the model are well consistent with the BCS theory. The second issue is about the property of the superfluid density on the QC. Our study clarifies that the pairing state obtained here exhibits the real SC that carries a finite superfluid density in the thermal dynamic limit and shows measurable Meissner effect. The insights acquired here also apply to other QC lattices.

\section*{Acknowledgement}
We are grateful to the stimulating discussions with Tao Li. This work is supported by the NSFC under the Grant Nos. 12074031, 12074037, 11674025, 11704029 and 11674151.

%\bibliographystyle{apsrev4-2}
%\bibliography{RefsForSCs.bib}

\begin{thebibliography}{36}%
\makeatletter
\providecommand \@ifxundefined [1]{%
 \@ifx{#1\undefined}
}%
\providecommand \@ifnum [1]{%
 \ifnum #1\expandafter \@firstoftwo
 \else \expandafter \@secondoftwo
 \fi
}%
\providecommand \@ifx [1]{%
 \ifx #1\expandafter \@firstoftwo
 \else \expandafter \@secondoftwo
 \fi
}%
\providecommand \natexlab [1]{#1}%
\providecommand \enquote  [1]{``#1''}%
\providecommand \bibnamefont  [1]{#1}%
\providecommand \bibfnamefont [1]{#1}%
\providecommand \citenamefont [1]{#1}%
\providecommand \href@noop [0]{\@secondoftwo}%
\providecommand \href [0]{\begingroup \@sanitize@url \@href}%
\providecommand \@href[1]{\@@startlink{#1}\@@href}%
\providecommand \@@href[1]{\endgroup#1\@@endlink}%
\providecommand \@sanitize@url [0]{\catcode `\\12\catcode `\$12\catcode
  `\&12\catcode `\#12\catcode `\^12\catcode `\_12\catcode `\%12\relax}%
\providecommand \@@startlink[1]{}%
\providecommand \@@endlink[0]{}%
\providecommand \url  [0]{\begingroup\@sanitize@url \@url }%
\providecommand \@url [1]{\endgroup\@href {#1}{\urlprefix }}%
\providecommand \urlprefix  [0]{URL }%
\providecommand \Eprint [0]{\href }%
\providecommand \doibase [0]{https://doi.org/}%
\providecommand \selectlanguage [0]{\@gobble}%
\providecommand \bibinfo  [0]{\@secondoftwo}%
\providecommand \bibfield  [0]{\@secondoftwo}%
\providecommand \translation [1]{[#1]}%
\providecommand \BibitemOpen [0]{}%
\providecommand \bibitemStop [0]{}%
\providecommand \bibitemNoStop [0]{.\EOS\space}%
\providecommand \EOS [0]{\spacefactor3000\relax}%
\providecommand \BibitemShut  [1]{\csname bibitem#1\endcsname}%
\let\auto@bib@innerbib\@empty
%</preamble>
\bibitem [{\citenamefont {Shechtman}\ \emph {et~al.}(1984)\citenamefont
  {Shechtman}, \citenamefont {Blech}, \citenamefont {Gratias},\ and\
  \citenamefont {Cahn}}]{Shechtman1951QCorder}%
  \BibitemOpen
  \bibfield  {author} {\bibinfo {author} {\bibfnamefont {D.}~\bibnamefont
  {Shechtman}}, \bibinfo {author} {\bibfnamefont {I.}~\bibnamefont {Blech}},
  \bibinfo {author} {\bibfnamefont {D.}~\bibnamefont {Gratias}},\ and\ \bibinfo
  {author} {\bibfnamefont {J.~W.}\ \bibnamefont {Cahn}},\ }\href
  {https://link.aps.org/doi/10.1103/PhysRevLett.53.1951} {\bibfield  {journal}
  {\bibinfo  {journal} {Phys. Rev. Lett.}\ }\textbf {\bibinfo {volume} {53}},\
  \bibinfo {pages} {1951} (\bibinfo {year} {1984})}\BibitemShut {NoStop}%
\bibitem [{\citenamefont {Goldman}\ and\ \citenamefont
  {Kelton}(1993)}]{Goldman1993QC}%
  \BibitemOpen
  \bibfield  {author} {\bibinfo {author} {\bibfnamefont {A.~I.}\ \bibnamefont
  {Goldman}}\ and\ \bibinfo {author} {\bibfnamefont {R.~F.}\ \bibnamefont
  {Kelton}},\ }\href {https://link.aps.org/doi/10.1103/RevModPhys.65.213}
  {\bibfield  {journal} {\bibinfo  {journal} {Rev. Mod. Phys.}\ }\textbf
  {\bibinfo {volume} {65}},\ \bibinfo {pages} {213} (\bibinfo {year}
  {1993})}\BibitemShut {NoStop}%
\bibitem [{\citenamefont {Senechal}(1996)}]{Senechal1996QC}%
  \BibitemOpen
  \bibfield  {author} {\bibinfo {author} {\bibfnamefont {M.}~\bibnamefont
  {Senechal}},\ }\href
  {https://www.cambridge.org/cn/academic/subjects/mathematics/mathematical-physics/quasicrystals-and-geometry?format=PB#contentsTabAnchor}
  {\emph {\bibinfo {title} {Quasicrystals and geometry}}}\ (\bibinfo
  {publisher} {ambridge University Press},\ \bibinfo {year} {1996})\BibitemShut
  {NoStop}%
\bibitem [{\citenamefont {Wessel}\ \emph {et~al.}(2003)\citenamefont {Wessel},
  \citenamefont {Jagannathan},\ and\ \citenamefont {Haas}}]{Wessel2003QC-QAF}%
  \BibitemOpen
  \bibfield  {author} {\bibinfo {author} {\bibfnamefont {S.}~\bibnamefont
  {Wessel}}, \bibinfo {author} {\bibfnamefont {A.}~\bibnamefont
  {Jagannathan}},\ and\ \bibinfo {author} {\bibfnamefont {S.}~\bibnamefont
  {Haas}},\ }\href {https://link.aps.org/doi/10.1103/PhysRevLett.90.177205}
  {\bibfield  {journal} {\bibinfo  {journal} {Phys. Rev. Lett.}\ }\textbf
  {\bibinfo {volume} {90}},\ \bibinfo {pages} {177205} (\bibinfo {year}
  {2003})}\BibitemShut {NoStop}%
\bibitem [{\citenamefont {Thiem}\ and\ \citenamefont
  {Chalker}(2015)}]{Thiem2015QC-QO}%
  \BibitemOpen
  \bibfield  {author} {\bibinfo {author} {\bibfnamefont {S.}~\bibnamefont
  {Thiem}}\ and\ \bibinfo {author} {\bibfnamefont {J.~T.}\ \bibnamefont
  {Chalker}},\ }\href {https://link.aps.org/doi/10.1103/PhysRevB.92.224409}
  {\bibfield  {journal} {\bibinfo  {journal} {Phys. Rev. B}\ }\textbf {\bibinfo
  {volume} {92}},\ \bibinfo {pages} {224409} (\bibinfo {year}
  {2015})}\BibitemShut {NoStop}%
\bibitem [{\citenamefont {Koga}\ and\ \citenamefont
  {Tsunetsugu}(2017)}]{Koga2017QC-AFO}%
  \BibitemOpen
  \bibfield  {author} {\bibinfo {author} {\bibfnamefont {A.}~\bibnamefont
  {Koga}}\ and\ \bibinfo {author} {\bibfnamefont {H.}~\bibnamefont
  {Tsunetsugu}},\ }\href {https://doi.org/10.1103/PhysRevB.96.214402}
  {\bibfield  {journal} {\bibinfo  {journal} {Phys. Rev. B}\ }\textbf {\bibinfo
  {volume} {96}},\ \bibinfo {pages} {214402} (\bibinfo {year}
  {2017})}\BibitemShut {NoStop}%
\bibitem [{\citenamefont {Shaginyan}\ \emph {et~al.}(2013)\citenamefont
  {Shaginyan}, \citenamefont {Msezane}, \citenamefont {Popov}, \citenamefont
  {Japaridze},\ and\ \citenamefont {Khodel}}]{Shaginyan2015QC-QPT}%
  \BibitemOpen
  \bibfield  {author} {\bibinfo {author} {\bibfnamefont {V.~R.}\ \bibnamefont
  {Shaginyan}}, \bibinfo {author} {\bibfnamefont {A.~Z.}\ \bibnamefont
  {Msezane}}, \bibinfo {author} {\bibfnamefont {K.~G.}\ \bibnamefont {Popov}},
  \bibinfo {author} {\bibfnamefont {G.~S.}\ \bibnamefont {Japaridze}},\ and\
  \bibinfo {author} {\bibfnamefont {V.~A.}\ \bibnamefont {Khodel}},\ }\href
  {https://doi.org/10.1103/PhysRevB.87.245122} {\bibfield  {journal} {\bibinfo
  {journal} {Phys. Rev. B}\ }\textbf {\bibinfo {volume} {87}},\ \bibinfo
  {pages} {245122} (\bibinfo {year} {2013})}\BibitemShut {NoStop}%
\bibitem [{\citenamefont {Otsuki}\ and\ \citenamefont
  {Kusunose}(2016)}]{Otsuki2016QC-QCB}%
  \BibitemOpen
  \bibfield  {author} {\bibinfo {author} {\bibfnamefont {J.}~\bibnamefont
  {Otsuki}}\ and\ \bibinfo {author} {\bibfnamefont {H.}~\bibnamefont
  {Kusunose}},\ }\href {https://doi.org/10.7566/JPSJ.85.073712} {\bibfield
  {journal} {\bibinfo  {journal} {J. Phys. Soc. Jap.}\ }\textbf {\bibinfo
  {volume} {85}},\ \bibinfo {pages} {073712} (\bibinfo {year}
  {2016})}\BibitemShut {NoStop}%
\bibitem [{\citenamefont {Watanabe}\ and\ \citenamefont
  {Miyake}(2016)}]{Watanabe2016QC-QCB}%
  \BibitemOpen
  \bibfield  {author} {\bibinfo {author} {\bibfnamefont {S.}~\bibnamefont
  {Watanabe}}\ and\ \bibinfo {author} {\bibfnamefont {K.}~\bibnamefont
  {Miyake}},\ }\href {https://doi.org/10.7566/JPSJ.85.063703} {\bibfield
  {journal} {\bibinfo  {journal} {J. Phys. Soc. Jap.}\ }\textbf {\bibinfo
  {volume} {85}},\ \bibinfo {pages} {063703} (\bibinfo {year}
  {2016})}\BibitemShut {NoStop}%
\bibitem [{\citenamefont {Takemori}\ and\ \citenamefont
  {Koga}(2015)}]{Takemori2015QC-QC}%
  \BibitemOpen
  \bibfield  {author} {\bibinfo {author} {\bibfnamefont {N.}~\bibnamefont
  {Takemori}}\ and\ \bibinfo {author} {\bibfnamefont {A.}~\bibnamefont
  {Koga}},\ }\href {https://doi.org/10.7566/JPSJ.84.023701} {\bibfield
  {journal} {\bibinfo  {journal} {J. Phys. Soc. Jap.}\ }\textbf {\bibinfo
  {volume} {84}},\ \bibinfo {pages} {023701} (\bibinfo {year}
  {2015})}\BibitemShut {NoStop}%
\bibitem [{\citenamefont {Takemura}\ \emph {et~al.}(2015)\citenamefont
  {Takemura}, \citenamefont {Takemori},\ and\ \citenamefont
  {Koga}}]{Takemura2015QC-QF}%
  \BibitemOpen
  \bibfield  {author} {\bibinfo {author} {\bibfnamefont {S.}~\bibnamefont
  {Takemura}}, \bibinfo {author} {\bibfnamefont {N.}~\bibnamefont {Takemori}},\
  and\ \bibinfo {author} {\bibfnamefont {A.}~\bibnamefont {Koga}},\ }\href
  {https://link.aps.org/doi/10.1103/PhysRevB.91.165114} {\bibfield  {journal}
  {\bibinfo  {journal} {Phys. Rev. B}\ }\textbf {\bibinfo {volume} {91}},\
  \bibinfo {pages} {165114} (\bibinfo {year} {2015})}\BibitemShut {NoStop}%
\bibitem [{\citenamefont {Andrade}\ \emph {et~al.}(2015)\citenamefont
  {Andrade}, \citenamefont {Jagannathan}, \citenamefont {Miranda},
  \citenamefont {Vojta},\ and\ \citenamefont
  {Dobrosavljevi\ifmmode~\acute{c}\else \'{c}\fi{}}}]{Andrade2015QC-QB}%
  \BibitemOpen
  \bibfield  {author} {\bibinfo {author} {\bibfnamefont {E.~C.}\ \bibnamefont
  {Andrade}}, \bibinfo {author} {\bibfnamefont {A.}~\bibnamefont
  {Jagannathan}}, \bibinfo {author} {\bibfnamefont {E.}~\bibnamefont
  {Miranda}}, \bibinfo {author} {\bibfnamefont {M.}~\bibnamefont {Vojta}},\
  and\ \bibinfo {author} {\bibfnamefont {V.}~\bibnamefont
  {Dobrosavljevi\ifmmode~\acute{c}\else \'{c}\fi{}}},\ }\href
  {https://link.aps.org/doi/10.1103/PhysRevLett.115.036403} {\bibfield
  {journal} {\bibinfo  {journal} {Phys. Rev. Lett.}\ }\textbf {\bibinfo
  {volume} {115}},\ \bibinfo {pages} {036403} (\bibinfo {year}
  {2015})}\BibitemShut {NoStop}%
\bibitem [{\citenamefont {Kraus}\ \emph {et~al.}(2012)\citenamefont {Kraus},
  \citenamefont {Lahini}, \citenamefont {Ringel}, \citenamefont {Verbin},\ and\
  \citenamefont {Zilberberg}}]{Kraus2012QC-TS}%
  \BibitemOpen
  \bibfield  {author} {\bibinfo {author} {\bibfnamefont {Y.~E.}\ \bibnamefont
  {Kraus}}, \bibinfo {author} {\bibfnamefont {Y.}~\bibnamefont {Lahini}},
  \bibinfo {author} {\bibfnamefont {Z.}~\bibnamefont {Ringel}}, \bibinfo
  {author} {\bibfnamefont {M.}~\bibnamefont {Verbin}},\ and\ \bibinfo {author}
  {\bibfnamefont {O.}~\bibnamefont {Zilberberg}},\ }\href
  {https://doi.org/10.1103/PhysRevLett.109.106402} {\bibfield  {journal}
  {\bibinfo  {journal} {Phys. Rev. Lett.}\ }\textbf {\bibinfo {volume} {109}},\
  \bibinfo {pages} {106402} (\bibinfo {year} {2012})}\BibitemShut {NoStop}%
\bibitem [{\citenamefont {Huang}\ and\ \citenamefont
  {Liu}(2018)}]{Huang2018QC-QSHS}%
  \BibitemOpen
  \bibfield  {author} {\bibinfo {author} {\bibfnamefont {H.}~\bibnamefont
  {Huang}}\ and\ \bibinfo {author} {\bibfnamefont {F.}~\bibnamefont {Liu}},\
  }\href {https://doi.org/10.1103/PhysRevLett.121.126401} {\bibfield  {journal}
  {\bibinfo  {journal} {Phys. Rev. Lett.}\ }\textbf {\bibinfo {volume} {121}},\
  \bibinfo {pages} {126401} (\bibinfo {year} {2018})}\BibitemShut {NoStop}%
\bibitem [{\citenamefont {Huang}\ and\ \citenamefont
  {Liu}(2019)}]{Huang2019QC-QSHS}%
  \BibitemOpen
  \bibfield  {author} {\bibinfo {author} {\bibfnamefont {H.}~\bibnamefont
  {Huang}}\ and\ \bibinfo {author} {\bibfnamefont {F.}~\bibnamefont {Liu}},\
  }\href {https://doi.org/10.1103/PhysRevB.100.085119} {\bibfield  {journal}
  {\bibinfo  {journal} {Phys. Rev. B}\ }\textbf {\bibinfo {volume} {100}},\
  \bibinfo {pages} {085119} (\bibinfo {year} {2019})}\BibitemShut {NoStop}%
\bibitem [{\citenamefont {Longhi}(2019)}]{Longhi2019QC-TPT}%
  \BibitemOpen
  \bibfield  {author} {\bibinfo {author} {\bibfnamefont {S.}~\bibnamefont
  {Longhi}},\ }\href {https://doi.org/10.1103/PhysRevLett.122.237601}
  {\bibfield  {journal} {\bibinfo  {journal} {Phys. Rev. Lett.}\ }\textbf
  {\bibinfo {volume} {122}},\ \bibinfo {pages} {237601} (\bibinfo {year}
  {2019})}\BibitemShut {NoStop}%
\bibitem [{\citenamefont {Wong}\ \emph {et~al.}(1987)\citenamefont {Wong},
  \citenamefont {Lopdrup}, \citenamefont {Wagner}, \citenamefont {Shen},\ and\
  \citenamefont {Poon}}]{Wong1987EXP}%
  \BibitemOpen
  \bibfield  {author} {\bibinfo {author} {\bibfnamefont {K.~M.}\ \bibnamefont
  {Wong}}, \bibinfo {author} {\bibfnamefont {E.}~\bibnamefont {Lopdrup}},
  \bibinfo {author} {\bibfnamefont {J.~L.}\ \bibnamefont {Wagner}}, \bibinfo
  {author} {\bibfnamefont {Y.}~\bibnamefont {Shen}},\ and\ \bibinfo {author}
  {\bibfnamefont {S.~J.}\ \bibnamefont {Poon}},\ }\href
  {https://link.aps.org/doi/10.1103/PhysRevB.35.2494} {\bibfield  {journal}
  {\bibinfo  {journal} {Phys. Rev. B}\ }\textbf {\bibinfo {volume} {35}},\
  \bibinfo {pages} {2494} (\bibinfo {year} {1987})}\BibitemShut {NoStop}%
\bibitem [{\citenamefont {Wagner}\ \emph {et~al.}(1988)\citenamefont {Wagner},
  \citenamefont {Biggs}, \citenamefont {Wong},\ and\ \citenamefont
  {Poon}}]{Wagner1988EXP}%
  \BibitemOpen
  \bibfield  {author} {\bibinfo {author} {\bibfnamefont {J.~L.}\ \bibnamefont
  {Wagner}}, \bibinfo {author} {\bibfnamefont {B.~D.}\ \bibnamefont {Biggs}},
  \bibinfo {author} {\bibfnamefont {K.~M.}\ \bibnamefont {Wong}},\ and\
  \bibinfo {author} {\bibfnamefont {S.~J.}\ \bibnamefont {Poon}},\ }\href
  {https://link.aps.org/doi/10.1103/PhysRevB.38.7436} {\bibfield  {journal}
  {\bibinfo  {journal} {Phys. Rev. B}\ }\textbf {\bibinfo {volume} {38}},\
  \bibinfo {pages} {7436} (\bibinfo {year} {1988})}\BibitemShut {NoStop}%
\bibitem [{\citenamefont {Deguchi}\ \emph {et~al.}(2015)\citenamefont
  {Deguchi}, \citenamefont {Nakayama}, \citenamefont {Matsukawa}, \citenamefont
  {Imura}, \citenamefont {Tanaka}, \citenamefont {Ishimasa},\ and\
  \citenamefont {Sato}}]{Deguchi2015EXP}%
  \BibitemOpen
  \bibfield  {author} {\bibinfo {author} {\bibfnamefont {K.}~\bibnamefont
  {Deguchi}}, \bibinfo {author} {\bibfnamefont {M.}~\bibnamefont {Nakayama}},
  \bibinfo {author} {\bibfnamefont {S.}~\bibnamefont {Matsukawa}}, \bibinfo
  {author} {\bibfnamefont {K.}~\bibnamefont {Imura}}, \bibinfo {author}
  {\bibfnamefont {K.}~\bibnamefont {Tanaka}}, \bibinfo {author} {\bibfnamefont
  {T.}~\bibnamefont {Ishimasa}},\ and\ \bibinfo {author} {\bibfnamefont
  {N.~K.}\ \bibnamefont {Sato}},\ }\href
  {https://doi.org/10.7566/JPSJ.84.023705} {\bibfield  {journal} {\bibinfo
  {journal} {Journal of the Physical Society of Japan}\ }\textbf {\bibinfo
  {volume} {84}},\ \bibinfo {pages} {023705} (\bibinfo {year}
  {2015})}\BibitemShut {NoStop}%
\bibitem [{\citenamefont {Kamiya}\ \emph {et~al.}(2018)\citenamefont {Kamiya},
  \citenamefont {Takeuchi}, \citenamefont {Kabeya}, \citenamefont {Wada},
  \citenamefont {Ishimasa}, \citenamefont {Ochiai}, \citenamefont {Deguchi},
  \citenamefont {Imura},\ and\ \citenamefont {Sato}}]{Kamiya2018qSC}%
  \BibitemOpen
  \bibfield  {author} {\bibinfo {author} {\bibfnamefont {K.}~\bibnamefont
  {Kamiya}}, \bibinfo {author} {\bibfnamefont {T.}~\bibnamefont {Takeuchi}},
  \bibinfo {author} {\bibfnamefont {N.}~\bibnamefont {Kabeya}}, \bibinfo
  {author} {\bibfnamefont {N.}~\bibnamefont {Wada}}, \bibinfo {author}
  {\bibfnamefont {T.}~\bibnamefont {Ishimasa}}, \bibinfo {author}
  {\bibfnamefont {A.}~\bibnamefont {Ochiai}}, \bibinfo {author} {\bibfnamefont
  {K.}~\bibnamefont {Deguchi}}, \bibinfo {author} {\bibfnamefont
  {K.}~\bibnamefont {Imura}},\ and\ \bibinfo {author} {\bibfnamefont {N.~K.}\
  \bibnamefont {Sato}},\ }\href {https://doi.org/10.1038/s41467-017-02667-x}
  {\bibfield  {journal} {\bibinfo  {journal} {Nat. Commun.}\ }\textbf {\bibinfo
  {volume} {9}},\ \bibinfo {pages} {154} (\bibinfo {year} {2018})}\BibitemShut
  {NoStop}%
\bibitem [{\citenamefont {Sakai}\ \emph {et~al.}(2017)\citenamefont {Sakai},
  \citenamefont {Takemori}, \citenamefont {Koga},\ and\ \citenamefont
  {Arita}}]{Sakai2017QC-SC}%
  \BibitemOpen
  \bibfield  {author} {\bibinfo {author} {\bibfnamefont {S.}~\bibnamefont
  {Sakai}}, \bibinfo {author} {\bibfnamefont {N.}~\bibnamefont {Takemori}},
  \bibinfo {author} {\bibfnamefont {A.}~\bibnamefont {Koga}},\ and\ \bibinfo
  {author} {\bibfnamefont {R.}~\bibnamefont {Arita}},\ }\href
  {https://link.aps.org/doi/10.1103/PhysRevB.95.024509} {\bibfield  {journal}
  {\bibinfo  {journal} {Phys. Rev. B}\ }\textbf {\bibinfo {volume} {95}},\
  \bibinfo {pages} {024509} (\bibinfo {year} {2017})}\BibitemShut {NoStop}%
\bibitem [{\citenamefont {Hou}\ \emph {et~al.}(2018)\citenamefont {Hou},
  \citenamefont {Hu}, \citenamefont {Sun},\ and\ \citenamefont
  {Zhang}}]{Hou2018Superfluid}%
  \BibitemOpen
  \bibfield  {author} {\bibinfo {author} {\bibfnamefont {J.}~\bibnamefont
  {Hou}}, \bibinfo {author} {\bibfnamefont {H.}~\bibnamefont {Hu}}, \bibinfo
  {author} {\bibfnamefont {K.}~\bibnamefont {Sun}},\ and\ \bibinfo {author}
  {\bibfnamefont {C.}~\bibnamefont {Zhang}},\ }\href
  {https://doi.org/10.1103/PhysRevLett.120.060407} {\bibfield  {journal}
  {\bibinfo  {journal} {Phys. Rev. Lett.}\ }\textbf {\bibinfo {volume} {120}},\
  \bibinfo {pages} {060407} (\bibinfo {year} {2018})}\BibitemShut {NoStop}%
\bibitem [{\citenamefont {Autti}\ \emph {et~al.}(2018)\citenamefont {Autti},
  \citenamefont {Eltsov},\ and\ \citenamefont {Volovik}}]{Autti2018Superfluid}%
  \BibitemOpen
  \bibfield  {author} {\bibinfo {author} {\bibfnamefont {S.}~\bibnamefont
  {Autti}}, \bibinfo {author} {\bibfnamefont {V.~B.}\ \bibnamefont {Eltsov}},\
  and\ \bibinfo {author} {\bibfnamefont {G.~E.}\ \bibnamefont {Volovik}},\
  }\href {https://doi.org/10.1103/PhysRevLett.120.215301} {\bibfield  {journal}
  {\bibinfo  {journal} {Phys. Rev. Lett.}\ }\textbf {\bibinfo {volume} {120}},\
  \bibinfo {pages} {215301} (\bibinfo {year} {2018})}\BibitemShut {NoStop}%
\bibitem [{\citenamefont {Ara\'ujo}\ and\ \citenamefont
  {Andrade}(2019)}]{Araujo2019QC-SC}%
  \BibitemOpen
  \bibfield  {author} {\bibinfo {author} {\bibfnamefont {R.~N.}\ \bibnamefont
  {Ara\'ujo}}\ and\ \bibinfo {author} {\bibfnamefont {E.~C.}\ \bibnamefont
  {Andrade}},\ }\href {https://doi.org/10.1103/PhysRevB.100.014510} {\bibfield
  {journal} {\bibinfo  {journal} {Phys. Rev. B}\ }\textbf {\bibinfo {volume}
  {100}},\ \bibinfo {pages} {014510} (\bibinfo {year} {2019})}\BibitemShut
  {NoStop}%
\bibitem [{\citenamefont {Sakai}\ and\ \citenamefont
  {Arita}(2019)}]{Sakai2019pairing}%
  \BibitemOpen
  \bibfield  {author} {\bibinfo {author} {\bibfnamefont {S.}~\bibnamefont
  {Sakai}}\ and\ \bibinfo {author} {\bibfnamefont {R.}~\bibnamefont {Arita}},\
  }\href {https://doi.org/10.1103/PhysRevResearch.1.022002} {\bibfield
  {journal} {\bibinfo  {journal} {Phys. Rev. Research}\ }\textbf {\bibinfo
  {volume} {1}},\ \bibinfo {pages} {022002} (\bibinfo {year}
  {2019})}\BibitemShut {NoStop}%
\bibitem [{\citenamefont {Takemori}\ \emph {et~al.}(2020)\citenamefont
  {Takemori}, \citenamefont {Arita},\ and\ \citenamefont
  {Sakai}}]{Takemori2020}%
  \BibitemOpen
  \bibfield  {author} {\bibinfo {author} {\bibfnamefont {N.}~\bibnamefont
  {Takemori}}, \bibinfo {author} {\bibfnamefont {R.}~\bibnamefont {Arita}},\
  and\ \bibinfo {author} {\bibfnamefont {S.}~\bibnamefont {Sakai}},\ }\href
  {https://doi.org/10.1103/PhysRevB.102.115108} {\bibfield  {journal} {\bibinfo
   {journal} {Phys. Rev. B}\ }\textbf {\bibinfo {volume} {102}},\ \bibinfo
  {pages} {115108} (\bibinfo {year} {2020})}\BibitemShut {NoStop}%
\bibitem [{\citenamefont {Nagai}(2020)}]{Nagai2020BdG}%
  \BibitemOpen
  \bibfield  {author} {\bibinfo {author} {\bibfnamefont {Y.}~\bibnamefont
  {Nagai}},\ }\href {https://doi.org/10.7566/JPSJ.89.074703} {\bibfield
  {journal} {\bibinfo  {journal} {J. Phys. Soc. Jap.}\ }\textbf {\bibinfo
  {volume} {89}},\ \bibinfo {pages} {074703} (\bibinfo {year}
  {2020})}\BibitemShut {NoStop}%
\bibitem [{\citenamefont {Cooper}(1956)}]{Cooper1956}%
  \BibitemOpen
  \bibfield  {author} {\bibinfo {author} {\bibfnamefont {L.~N.}\ \bibnamefont
  {Cooper}},\ }\href {https://doi.org/10.1103/PhysRev.104.1189} {\bibfield
  {journal} {\bibinfo  {journal} {Phys. Rev.}\ }\textbf {\bibinfo {volume}
  {104}},\ \bibinfo {pages} {1189} (\bibinfo {year} {1956})}\BibitemShut
  {NoStop}%
\bibitem [{\citenamefont {Bardeen}\ \emph {et~al.}(1957)\citenamefont
  {Bardeen}, \citenamefont {Cooper},\ and\ \citenamefont
  {Schrieffer}}]{Bardeen1957Theory}%
  \BibitemOpen
  \bibfield  {author} {\bibinfo {author} {\bibfnamefont {J.}~\bibnamefont
  {Bardeen}}, \bibinfo {author} {\bibfnamefont {L.~N.}\ \bibnamefont
  {Cooper}},\ and\ \bibinfo {author} {\bibfnamefont {J.~R.}\ \bibnamefont
  {Schrieffer}},\ }\href {https://link.aps.org/doi/10.1103/PhysRev.108.1175}
  {\bibfield  {journal} {\bibinfo  {journal} {Phys. Rev.}\ }\textbf {\bibinfo
  {volume} {108}},\ \bibinfo {pages} {1175} (\bibinfo {year}
  {1957})}\BibitemShut {NoStop}%
\bibitem [{\citenamefont {Schrieffer}(2018)}]{SchriefferBook}%
  \BibitemOpen
  \bibfield  {author} {\bibinfo {author} {\bibfnamefont {J.~R.}\ \bibnamefont
  {Schrieffer}},\ }\href@noop {} {\emph {\bibinfo {title} {Theory Of
  Superconductivity}}}\ (\bibinfo  {publisher} {CRC press},\ \bibinfo {address}
  {Boca Raton, FL},\ \bibinfo {year} {2018})\BibitemShut {NoStop}%
\bibitem [{\citenamefont {Tsunetsugu}\ and\ \citenamefont
  {Ueda}(1991)}]{Tsunetsugu1991Conductance}%
  \BibitemOpen
  \bibfield  {author} {\bibinfo {author} {\bibfnamefont {H.}~\bibnamefont
  {Tsunetsugu}}\ and\ \bibinfo {author} {\bibfnamefont {K.}~\bibnamefont
  {Ueda}},\ }\href {https://doi.org/10.1103/PhysRevB.43.8892} {\bibfield
  {journal} {\bibinfo  {journal} {Phys. Rev. B}\ }\textbf {\bibinfo {volume}
  {43}},\ \bibinfo {pages} {8892} (\bibinfo {year} {1991})}\BibitemShut
  {NoStop}%
\bibitem [{\citenamefont {Anderson}(1959)}]{Anderson}%
  \BibitemOpen
  \bibfield  {author} {\bibinfo {author} {\bibfnamefont {P.}~\bibnamefont
  {Anderson}},\ }\href
  {https://doi.org/https://doi.org/10.1016/0022-3697(59)90036-8} {\bibfield
  {journal} {\bibinfo  {journal} {Journal of Physics and Chemistry of Solids}\
  }\textbf {\bibinfo {volume} {11}},\ \bibinfo {pages} {26 } (\bibinfo {year}
  {1959})}\BibitemShut {NoStop}%
\bibitem [{\citenamefont {Tsunetsugu}\ \emph {et~al.}(1991)\citenamefont
  {Tsunetsugu}, \citenamefont {Fujiwara}, \citenamefont {Ueda},\ and\
  \citenamefont {Tokihiro}}]{Tsunetsugu1991Wavefunction}%
  \BibitemOpen
  \bibfield  {author} {\bibinfo {author} {\bibfnamefont {H.}~\bibnamefont
  {Tsunetsugu}}, \bibinfo {author} {\bibfnamefont {T.}~\bibnamefont
  {Fujiwara}}, \bibinfo {author} {\bibfnamefont {K.}~\bibnamefont {Ueda}},\
  and\ \bibinfo {author} {\bibfnamefont {T.}~\bibnamefont {Tokihiro}},\ }\href
  {https://doi.org/10.1103/PhysRevB.43.8879} {\bibfield  {journal} {\bibinfo
  {journal} {Phys. Rev. B}\ }\textbf {\bibinfo {volume} {43}},\ \bibinfo
  {pages} {8879} (\bibinfo {year} {1991})}\BibitemShut {NoStop}%
\bibitem [{\citenamefont {Yamamoto}\ and\ \citenamefont
  {Fujiwara}(1995)}]{Yamamoto1995}%
  \BibitemOpen
  \bibfield  {author} {\bibinfo {author} {\bibfnamefont {S.}~\bibnamefont
  {Yamamoto}}\ and\ \bibinfo {author} {\bibfnamefont {T.}~\bibnamefont
  {Fujiwara}},\ }\href {https://doi.org/10.1103/PhysRevB.51.8841} {\bibfield
  {journal} {\bibinfo  {journal} {Phys. Rev. B}\ }\textbf {\bibinfo {volume}
  {51}},\ \bibinfo {pages} {8841} (\bibinfo {year} {1995})}\BibitemShut
  {NoStop}%
\bibitem [{\citenamefont {Cao}\ \emph {et~al.}(2020)\citenamefont {Cao},
  \citenamefont {Zhang}, \citenamefont {Liu}, \citenamefont {Liu},
  \citenamefont {Chen},\ and\ \citenamefont {Yang}}]{Cao2020QCSC}%
  \BibitemOpen
  \bibfield  {author} {\bibinfo {author} {\bibfnamefont {Y.}~\bibnamefont
  {Cao}}, \bibinfo {author} {\bibfnamefont {Y.}~\bibnamefont {Zhang}}, \bibinfo
  {author} {\bibfnamefont {Y.-B.}\ \bibnamefont {Liu}}, \bibinfo {author}
  {\bibfnamefont {C.-C.}\ \bibnamefont {Liu}}, \bibinfo {author} {\bibfnamefont
  {W.-Q.}\ \bibnamefont {Chen}},\ and\ \bibinfo {author} {\bibfnamefont
  {F.}~\bibnamefont {Yang}},\ }\href
  {https://doi.org/10.1103/PhysRevLett.125.017002} {\bibfield  {journal}
  {\bibinfo  {journal} {Phys. Rev. Lett.}\ }\textbf {\bibinfo {volume} {125}},\
  \bibinfo {pages} {017002} (\bibinfo {year} {2020})}\BibitemShut {NoStop}%
\bibitem [{\citenamefont {Mahan}(2000)}]{MahanBook}%
  \BibitemOpen
  \bibfield  {author} {\bibinfo {author} {\bibfnamefont {J.~D.}\ \bibnamefont
  {Mahan}},\ }\href@noop {} {\emph {\bibinfo {title} {Many-Particle Physics}}}\
  (\bibinfo  {publisher} {Springer, Boston, MA},\ \bibinfo {year}
  {2000})\BibitemShut {NoStop}%
\end{thebibliography}
%apsrev4-2.bst 2019-01-14 (MD) hand-edited version of apsrev4-1.bst
%Control: key (0)
%Control: author (72) initials jnrlst
%Control: editor formatted (1) identically to author
%Control: production of article title (-1) disabled
%Control: page (0) single
%Control: year (1) truncated
%Control: production of eprint (0) enabled
%

\end{document}